\documentclass[runningheads]{llncs}
\usepackage{graphicx} 
\usepackage{xcolor}
\usepackage{microtype}
\usepackage{amsfonts}
\usepackage{cite}
\usepackage{amsmath,amssymb,amsfonts}
\usepackage{mathtools}
\usepackage{mdframed}
\usepackage{enumitem}
\usepackage[colorlinks,linkcolor={blue!50!black},
    citecolor={blue!50!black},
    urlcolor={blue!80!black}]{hyperref}
\usepackage{tikz}
\usepackage{listings}
\usepackage{caption}
\usepackage{wrapfig}
\usepackage{subcaption}
\usepackage{multirow}
\usepackage[algoruled, linesnumbered, noend]{algorithm2e}
\SetKwIF{IfNot}{ElseIfNot}{}{if not}{then}{else if not}{}{}

\SetCommentSty{mycommfont}

\usepackage{booktabs,siunitx}

\usetikzlibrary{backgrounds,patterns,patterns.meta,arrows,arrows.meta,calc,shapes,shadows,decorations.pathmorphing,decorations.pathreplacing,automata,shapes.multipart,positioning,shapes.geometric,fit,circuits,trees,shapes.gates.logic.US,fit, shadows.blur,shapes.symbols,intersections}

\newcommand{\mathsymbol}[2]{ \newcommand{#1}{\ensuremath{#2}\xspace} }

\newcommand{\unitinterval}{[0,1]}

\newcommand{\sinit}{s_{0}}
\newcommand{\Act}{Act}
\newcommand{\act}{\ensuremath{\alpha}}
\newcommand{\mpm}{P}

\newcommand{\mdpT}{(S,\sinit,Act,\mpm)}
\newcommand{\mcT}{(S,\sinit,\mpm)}

\newcommand{\F}[1]{\Diamond #1}

\newcommand{\prob}[2][]{\mathbb{P}\left[ #1 \models #2\right ]}
\newcommand{\probmax}[2][]{\mathbb{P}_{\max}\left[ #1 \models #2\right ]}
\newcommand{\probmin}[2][]{\mathbb{P}_{\min}\left[ #1 \models #2\right ]}
\newcommand{\probthreshold}[3][]{\mathbb{P}_{#2}\left[ #1 \models #3 \right]}
\newcommand{\playeronemax}[2][]{\mathbb{P}_{\max}\left[ #1 \models #2\right ]}

\newcommand{\reach}[1]{\ensuremath{ \prob[#1]{\F{T}} }}
\newcommand{\reachmax}[1]{\ensuremath{ \probmax[#1]{\F{T}} }}

\newcommand{\sched}[1][]{\sigma_{#1}}
\newcommand{\schedrandom}[1][]{\sigma_{\textrm{rand}}}

\mathsymbol{\qsched}{\sched[\qmdp]}
\mathsymbol{\gsched}{\sched[\game]}

\mathsymbol{\unsat}{\varnothing}
\mathsymbol{\schedulers}{\Sigma^M}
\mathsymbol{\fschedulers}{\Sigma^{\family}}

\mathsymbol{\indices}{\mathcal{I}}
\newcommand{\indexedmdps}[1][]{\{M_i\}_{i #1}}

\newcommand{\widesim}[1][1.5]{ \mathrel{ \scalebox{#1}[1]{$\sim$} } }
\mathsymbol{\similarto}{ \,\substack{s,\act \\ \widesim[1.5] }\,}

\mathsymbol{\indicespartition}{\indices /\! \similarto}
\mathsymbol{\subindices}{I}

\mathsymbol{\family}{\mathcal{M}}
\mathsymbol{\states}{S_{\family}}
\mathsymbol{\actions}{Act_{\family}}

\mathsymbol{\bmp}{\mathcal{B}}
\mathsymbol{\pmp}{\mathcal{P}}

\mathsymbol{\qmdp}{\mathcal{Q}}
\mathsymbol{\coloring}{\Gamma}
\mathsymbol{\qmdpT}{ \left(\qmdp,\coloring\right) }

\mathsymbol{\ptree}{\mathcal{T}}
\mathsymbol{\nodelabel}{F}
\mathsymbol{\leaflabel}{L}

\mathsymbol{\game}{\mathcal{G}}

\mathsymbol{\reachable}{\mathsf{Reach}}
\mathsymbol{\consdindices}{\mathsf{ConsIDs}}

\mathsymbol{\indicesone}{\indices_1}
\mathsymbol{\indicestwo}{\indices_2}
\mathsymbol{\familyone}{\family_1}
\mathsymbol{\familytwo}{\family_2}

\newcommand{\tool}[1]{{\textsc{#1}}}
\newcommand{\storm}{\tool{Storm}\xspace}
\newcommand{\paynt}{\tool{Paynt}\xspace}

\def\orcidID#1{\smash{\href{http://orcid.org/#1}{\protect\raisebox{-1.25pt}{\protect\includegraphics{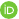}}}}}

\renewcommand{\paragraph}[1]{\smallskip\noindent\emph{#1}}
\renewcommand{\subsubsection}[1]{\medskip\noindent\textbf{#1}}

\definecolor{prismgreen}{HTML}{009900}
\definecolor{prismred}{HTML}{cc0000}
\definecolor{prismblue}{HTML}{0000cc}

\lstdefinelanguage{Prism}{
        basicstyle=\color{prismred}\scriptsize\ttfamily,
        literate=*	{0}{{\textcolor{prismblue}{0}}}{1}
			{1}{{\textcolor{prismblue}{1}}}{1}
			{2}{{\textcolor{prismblue}{2}}}{1}
			{3}{{\textcolor{prismblue}{3}}}{1}
			{4}{{\textcolor{prismblue}{4}}}{1}
			{5}{{\textcolor{prismblue}{5}}}{1}
			{6}{{\textcolor{prismblue}{6}}}{1}
			{7}{{\textcolor{prismblue}{7}}}{1}
			{8}{{\textcolor{prismblue}{8}}}{1}
			{9}{{\textcolor{prismblue}{9}}}{1}
			{.0}{{\textcolor{prismblue}{.0}}}{2}
			{.1}{{\textcolor{prismblue}{.1}}}{2}
			{.2}{{\textcolor{prismblue}{.2}}}{2}
			{.3}{{\textcolor{prismblue}{.3}}}{2}
			{.4}{{\textcolor{prismblue}{.4}}}{2}
			{.5}{{\textcolor{prismblue}{.5}}}{2}
			{.6}{{\textcolor{prismblue}{.6}}}{2}
			{.7}{{\textcolor{prismblue}{.7}}}{2}
			{.8}{{\textcolor{prismblue}{.8}}}{2}
			{.9}{{\textcolor{prismblue}{.9}}}{2}
			{[}{{\textcolor{black}{[}}}{1}
			{]}{{\textcolor{black}{]}}}{1}
			{+}{{\textcolor{black}{+}}}{1}
			{-}{{\textcolor{black}{-}}}{1}
			{=}{{\textcolor{black}{=}}}{1}
			{<}{{\textcolor{black}{<}}}{1}
			{>}{{\textcolor{black}{>}}}{1}
			{\&}{{\textcolor{black}{\&}}}{1}
			{|}{{\textcolor{black}{|}}}{1}
			{:}{{\textcolor{black}{:}}}{1}
			{;}{{\textcolor{black}{;}}}{1}
			{(}{{\textcolor{black}{(}}}{1}
			{)}{{\textcolor{black}{)}}}{1}
			{..}{{\textcolor{black}{..}}}{2},
        keywords= {bool,ceil,const,ctmc,double,dtmc,endinit,endmodule,endrewards, endsystem,F,false,floor,formula,G,global,I,init,int,label,max,mdp,min,module,nondeterministic,P,Pmin,Pmax,prob,probabilistic,rate,rewards,Rmin,Rmax,S,stochastic,system,true,U, option, either, assignment, relation, operation, hole, variable},
        keywordstyle={\bfseries\color{black}},
        numberstyle=\footnotesize\color{black},
        comment=[l] {//}, morecomment=[s]{/*}{*/},
        commentstyle= \color{prismgreen},
        tabsize=4,
        captionpos=b,
        escapechar=^,
        moredelim=[is][\color{orange}]{@}{@},
}

\title{Policies Grow on Trees:\\Model Checking Families of MDPs\thanks{This work is partially supported by the Czech Science Foundation grant \mbox{GA23-06963S} (VESCAA) the NWO Veni grant ProMiSe (222.147).}}

\author{}
\institute{}
 \author{Roman Andriushchenko\inst{1}\orcidID{0000-0002-1286-934X} \and  Milan \v{C}e\v{s}ka\inst{1}\orcidID{0000-0002-0300-9727} \and Sebastian Junges\inst{2}\orcidID{0000-0003-0978-8466}  \and Filip Mac\'{a}k\inst{1}\orcidID{0009-0004-4277-2751}}
\institute{Brno University of Technology, Brno, Czech Republic \and Radboud University, Nijmegen, the Netherlands}

\begin{document}

\maketitle

\begin{abstract}
Markov decision processes (MDPs) provide a fundamental model for sequential decision making under process uncertainty. A classical synthesis task is to compute for a given MDP a winning policy that achieves a desired specification. However, at design time, one typically needs to consider a family of MDPs modelling various system variations. For a given family, we study synthesising (1)~the subset of MDPs where a winning policy exists and (2)~a preferably small number of winning policies that together cover this subset. We introduce policy trees that concisely capture the synthesis result. The key ingredient for synthesising policy trees is a recursive application of a game-based abstraction. We combine this abstraction with an efficient refinement procedure and a post-processing step. An extensive empirical evaluation demonstrates superior scalability of our approach compared to naive baselines. For one of the benchmarks, we find 246~winning policies covering 94~million MDPs. Our algorithm requires less than 30 minutes, whereas the naive baseline only covers 3.7\% of MDPs in 24 hours.
\end{abstract}


\section{Introduction}

Markov decision processes (MDPs) are the ubiquitous model to describe sequential decision making uncertainty: the outcomes of nondeterministic actions are determined by a probability distribution over the successor states. MDP policies resolve the non-determinism and describe for each state which action to take. A classical synthesis task in MDPs is to compute policies that achieve reachability specifications: Given a set of target states and a threshold $\lambda$, find a \emph{winning policy}, i.e., a policy that ensures that the targets are reached with probability at least $\lambda$, if one exists.  Winning policies are efficiently computed by modern probabilistic model checkers such as~Storm~\cite{STORM} or Prism~\cite{KNP11}. 

In this paper, we generalise the synthesis task: We consider not one, but a set of MDPs. The MDPs model variations of the same system, this is a natural approach to capture alternative environment conditions for an agent~\cite{DBLP:conf/aaai/ChadesCMNSB12}, protocol variations~\cite{DBLP:journals/infsof/GhezziS13}, or alternative hardware choices~\cite{DBLP:conf/nfm/0001HTT19}. In such scenarios, even huge sets of MDPs can be concisely represented. 
The goal is to compute a (potentially different) winning policy for every MDP, if possible. Naively, this problem is straightforward to solve using either of two baselines: Either we \emph{enumerate} over the MDPs and solve each MDP individually, or we create an \emph{all-in-one} MDP that contains the disjoint union of all MDPs and obtain a policy for that (giant) MDP~\cite{allinone,DBLP:conf/hase/RodriguesANLCSS15}. The first challenge that this paper tackles is \emph{to support solving this problem for families with millions of MDPs}, where both baselines have clear scalability issues\footnote{The all-in-one MDP may be stored symbolically~\cite{DBLP:journals/fac/ChrszonDKB18}, however, in many cases, existing symbolic techniques for probabilistic model checking do not work well~\cite{budde2020correctness}.}.  Furthermore, we consider the representation of the resulting set of policies. We conjecture that a small number of different policies is often sufficient to win in every MDP. Thus, as a second challenge, \emph{we want to find a concise representation of a function that maps from the MDP to a winning policy}.

\begin{figure}[t]
    \centering
    \begin{subfigure}[t]{0.45\textwidth}
        \centering
        \includegraphics[width=0.85\textwidth]{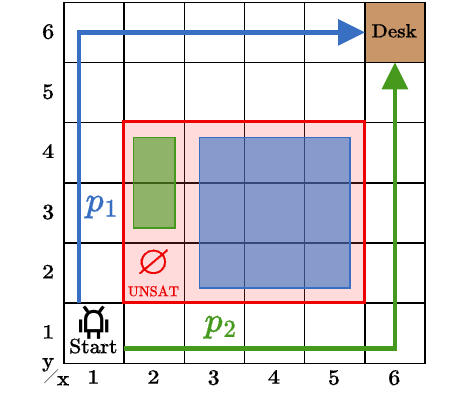}
        \caption{ The robot wants to reach the desk safely. The red zone shows potential locations for the obstacle. The slippiness is given in (c) to the right.}
        \label{fig:running-example:grid}
    \end{subfigure}~
    \begin{subfigure}[t]{0.49\textwidth}
        \centering
        \includegraphics[width=0.9\textwidth]{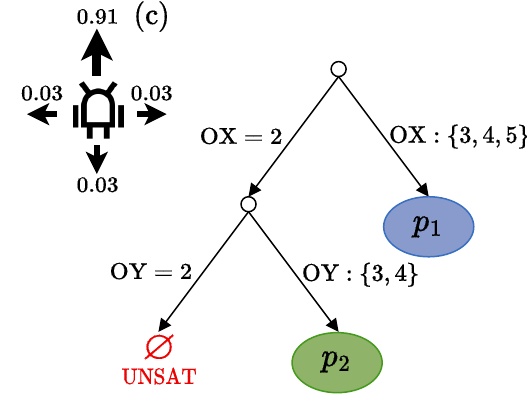}
        \caption{Policy tree generated by our approach, with edges labelled by propositions on chair positions (OX, OY). Leaves contain a  winning policy or \unsat, if no winning policy exists.}
        \label{fig:running-example:policies}
    \end{subfigure}
    \caption{Illustrative example for a set of MDPs and a policy tree}
    \label{fig:running-example}
    \vspace{-1em}
\end{figure}

\paragraph{Illustrative example.} We consider a delivery robot that brings coffee to a desk. Rooms are parameterised by the location of obstacles such as chairs, in this case of a single chair at (OX, OY). Upon entering, the robot scans the room to obtain the location of the obstacles and plans how to reach the desk. We do not consider exiting the room afterwards. Due to the imprecision of the actuators, the system can be modelled as a slippery gridworld. See Fig.~\ref{fig:running-example:grid}. As a manufacturer, we aim to ensure a 99$\%$ success rate for various room configurations. The policy that ensures this success rate may depend on the location of the chair. We want to obtain a solution that looks as illustrated in Fig.~\ref{fig:running-example:policies}: If a chair is in the upper left quadrant, our policy travels via the right bottom corner; if the chair is on the right,  the robot travels via the left wall. If the chair is near the entrance, no policy ensures reaching the desk with more than $99\%$ probability. 


\paragraph{Problem setup.}
Let us clarify the setup of the relevant problems.
We have indexed sets of MDPs, which are represented using any program-like description language\footnote{We use the standard Prism language~\cite{KNP11}.} featuring some undefined integer constants or input arguments. The assignments to these constants can be used as indices for the individual MDPs: In our example, we have an MDP for $(\textrm{OX}{=}3,\textrm{OY}{=}3)$, an MDP for $(\textrm{OX}{=}4,\textrm{OY}{=}2)$, etc. For a fixed description, a set of indices corresponds to a set of MDPs.
For each MDP, we want to point to a potentially different winning policy. We suggest using \emph{policy trees}: These trees recursively partition the set of MDPs into subsets. The leaves of the tree thus correspond to a set of MDPs. If such a leaf is labelled with a policy $\sched$, this policy is winning for all MDPs in the subset. Leafs labelled with $\unsat$ indicate no MDP in the subset has any winning policy. 


\paragraph{One policy or different policies.}
As we search for a policy for every MDP individually, it suffices to assume that winning policies are memoryless, i.e., mappings from states to actions. Furthermore, in this paper, we assume an arbitrary but atomic\footnote{dtControl~\cite{DBLP:conf/tacas/AshokJKWWY21} or Uppaal~\cite{DBLP:conf/qest/AshokKLCTW19} concisely represent individual policies, e.g., as decision trees. We are not considering the space requirement of representing the policies. } representation of these policies. However, our leaves are labelled with a single policy that is winning on all MDPs. The question of whether such a \emph{robust} policy exists is, in general, much harder~\cite{arming2018parameter}, since it can profit from (arbitrary) memory: Intuitively, based on the observed transitions, the policy may infer the likelihood of being applied to a particular MDP. The problem is indeed closely related to solving policies for partially observable MDPs. We avoid the computational costs and the potential lack of concise policies by considering only memoryless policies. However, even for an explicitly represented set of MDPs, it is NP-complete to consider the existence of such a policy. This is in contrast to the polynomial-time complexity of finding winning policies for every MDP.







\paragraph{A game-based approach.}
To find a (concise) expression capturing all MDPs that allow for a winning policy, 
we propose an adaptation of a game-based abstraction approach~\cite{DBLP:conf/tacas/HahnHWZ10,DBLP:journals/fmsd/KattenbeltKNP10} that has been successful in the verification of sets of Markov chains~\cite{DBLP:conf/cav/AndriushchenkoC21} and for parametric MDPs~\cite{DBLP:journals/corr/abs-1903-07993}. We discuss differences in Section~\ref{sec:related}. 
In a nutshell, the game-based approach operates as follows: from the concise representation of the set of MDPs, a stochastic game (SG) is constructed that mimics all MDPs at once. One player selects actions in the MDP, and the other player selects in which MDP to evaluate the action. The constructed game is an abstraction, as the player choosing the MDPs can switch between MDPs based on the current state of the system. 
If the SG satisfies a reachability property, then there is a winning policy for every MDP. In fact, there must be a robust winning policy, and this robust policy is easy to obtain. 
The reverse does not hold. Thus, we either find a robust policy or obtain an inconclusive result. In the latter case, we divide-and-conquer: We split the set of MDPs and recursively invoke our algorithm on the smaller sets. The abstraction is effective when the number of necessary splits remains limited. Importantly, the size of the resulting tree corresponds to the number of~splits.


\paragraph{Practical efficacy.}
An abstraction-refinement approach must be paired with efficient heuristics for refinement. We carefully study several variations that are guided by the model checking results. 
Furthermore, once we have found robust policies for various subfamilies, they are a good basis for further pruning of the tree, thereby yielding smaller trees with negligible overhead.

\paragraph{Alternatives.}
We studied a rich set of alternatives to the main approach we present in the paper. 
Most prominently, to ensure that our game-based abstraction is not overly conservative, we compute policies that work well against a (uniform) distribution and then only verify whether they are robust. Our experiments show that the game-based abstraction dominates such an alternative. 

\paragraph{Contributions and structure.}
In summary, we contribute a scalable approach to policy synthesis for sets of MDPs. The key technique is a game-based abstraction (Sec.~\ref{sec:AR}), embedded in a recursive approach with context-specific ways to refine the abstraction (Sec.~\ref{sec:algorithm}). The result is an algorithm that finds policies for millions of (similar) MDPs at once and represents the necessary policies compactly. Our extensive evaluation demonstrates (Sec.~\ref{sec:experiments}) the superior scalability and discusses strengths and weaknesses, along with an ablation study to understand the origins of the performance. We extensively discuss further related work in Sec.~\ref{sec:related}.

\section{Problem Statement}

\subsubsection{Preliminaries}
A \emph{distribution} over a countable set $A$ is a~function $\mu \colon A \rightarrow \unitinterval$ s.t.~$\sum_a \mu(a) {=} 1$.
The set $Distr(A)$ contains all distributions over $A$.

\begin{definition}[MDP]
A \emph{Markov decision process (MDP)} is a tuple $M = \mdpT$ with a countable set $S$ of states, an initial state $\sinit \in S$, a finite set $\Act$ of actions, and a partial transition function $\mpm \colon S \times \Act \nrightarrow Distr(S)$.
\end{definition}
For an MDP $M$, we define the actions available in  $s \in S$ as
$\Act^M(s) \coloneqq \left\{ \act \in \Act \mid \mpm(s,\act) \neq \bot \right\}$. We omit $M$ whenever it is clear from the context.
We assume $\Act(s) \neq \emptyset$ for each $s \in S$ (no deadlocks).
An MDP with $|\Act(s)|=1$ for each $s \in S$ is a \emph{Markov chain (MC)}.
We denote MCs as a tuple $\mcT$.
We denote $\mpm(s,\act,s') \coloneqq \mpm(s,\act)(s')$.
A (finite) \emph{path} of an MDP $M$ is a sequence $\pi = s_0\act_0s_1\act_1\dots s_n$ where ${\mpm(s_i,\act_i,s_{i+1}) > 0}$ for $0 {\leq} i {<} n$. 
A (deterministic, memoryless) \emph{policy} is a function $\sigma \colon S \rightarrow \Act$ where $\sigma(s) \in \Act(s)$ for all $s \in S$. The set $\schedulers$ denotes the policies for MDP $M$.
A~policy $\sigma \in \schedulers$ induces the MC~$M^{\sigma} = \left(S,s_0,\mpm^\sigma\right)$ where \mbox{$\mpm^\sigma(s) = \mpm(s,\sigma(s))$}.

\paragraph{Specifications and winning policies.} We consider indefinite-horizon reachability properties~\cite{puterman2014markov}.  Formally, let $M=\mcT$ be an MC, and let $T \subseteq S$ be a set of \emph{target states}.
The probability of a finite path is the product of the individual transition probabilities. 
$\reach{M}$ denotes the probability of reaching~$T$ in~$M$ from the initial state $\sinit$, corresponding to the integral over the probabilities of all paths ending in $T$.
Now assume MDP $M = \mdpT$.
The \emph{maximum reachability probability} of $T$ from state in $M$ is $\reachmax{M} \coloneqq \sup_{\sigma \in \schedulers} \reach{M^\sigma}$.
In the remainder of the paper, we consider specifications of type $\varphi = \probthreshold[M]{\geq \lambda}{\F{T}}$ for a given threshold $\lambda \in [0,1]$.
We say that policy~$\sigma$ is \emph{winning} for MDP $M$ iff $\reach{M^\sigma} \geq \lambda$;
We also write $M, \sigma \models \varphi$.
We say that MDP $M$ is \emph{satisfiable} if there exists a winning policy for $M$.
Formal definitions can be found, e.g. in~\cite{Baier2018}.

\vspace{-.5em}
\subsection{Families of MDPs}
\label{sec:familymdp}
We consider indexed sets of MDPs that share a common state and action space.
\begin{definition}[Family of MDPs]
A \emph{family of MDPs} over the set $\states$ of states and set $\actions$ of actions is an indexed set $\family = \{ (\states,\sinit,\actions,\mpm_i) \}_{i \in \indices}$ of MDPs, where $\indices$ is a finite index set of \emph{identifiers}.
We assume that for each state, the sets of available actions coincide in all MDPs: $\forall s \in \states \colon \forall i,j \in \indices \colon \mpm_i(s,\act) \neq \bot \Rightarrow \mpm_j(s,\act) \neq \bot$.
As a consequence, all MDPs in the family have the same set of available policies, which will be denoted as \fschedulers.
\end{definition}
Importantly, the MDPs differ in their transition functions and therefore may have different reachable state spaces.
Note that the shape of the index set $\indices$ is not important for the paper, but it is crucial for the implementation of some heuristics. For instance, the family from Fig.~\ref{fig:running-example} is indexed using set $\{({\textrm{OX}{=}2},\textrm{OY}{=}2),(\textrm{OX}{=}2,\textrm{OY}{=}3),\dots\}$.
Let $\similarto$ be the equivalence relation on \indices defined as $i \similarto j$ iff $\mpm_i(s,\act) = \mpm_j(s,\act)$, i.e.~MDPs $M_i$ and $M_j$ execute the same action $\act$ in state $s$. \indicespartition denotes the corresponding equivalence partitioning of \indices wrt.~$\similarto$. We will use \indicespartition later to efficiently encode families of MDPs. An \emph{all-in-one MDP} is simply the disjoint union of the family members. Conceptually, it has as many initial states as there are MDPs in the family\footnote{
A dummy initial state having $n$ actions to the original initial states can be added.}.

For MDP families, we distinguish two different problems below.

\paragraph{1. Policy map.}
We are interested in finding winning policies for every MDP in a family. However, a family may also contain MDPs for which no winning policy exists. 
Satisficing policy maps take this into account:

\begin{definition}[Policy map]
Given a family of MDPs \family and a specification~$\varphi$. A \emph{(satisficing) policy map} is a function $\pmp\colon \family \rightarrow \fschedulers \cup \{\unsat\}$ such that $\pmp(M_i) = \sigma$ if $M_i,\sigma \models \varphi$, and $\pmp(M_i) = \unsat$ if
$\forall  \sched{} \in \fschedulers, M_i,\sigma \not\models \varphi.$ 

\end{definition}
Satisficing policy maps can be computed in time linear in the number of family members and polynomial in the size of the MDPs in that family\footnote{The complexity depends on the specification; this indeed holds for reachability.}. In particular, a simple enumerative algorithm invoking a tractable MDP model checking routine can achieve this. Alternatively, a tractable algorithm can be obtained by model checking the all-in-one MDP.

\paragraph{2. Robust policy.}
Robust policies are single policies that are winning on a set of MDPs. In general, robust policies may require memory; however, this paper is not focused on finding general robust policies. Therefore, we can restrict our formal definition to memoryless policies.
\begin{definition}[Robust policy]
Given a family of MDPs $\family$ and a specification $\varphi$. A policy $\sched{} \in \fschedulers$ is called a \emph{robust policy} if $\forall M_i \in \family: M_i,\sigma \models \varphi$.
\end{definition}

\begin{figure}[t]

    \begin{subfigure}[b]{.33\textwidth}
    \centering
    \includegraphics[width=0.78\textwidth]{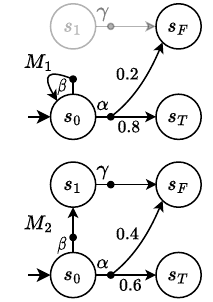}
    \caption{}

    \label{fig:example:mdp-family}
    \end{subfigure}%
    \begin{subfigure}{.33\textwidth}
    \centering
    \includegraphics[width=0.78\textwidth]{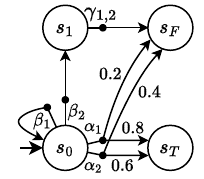}
    \caption{}
    \label{fig:example:quotient}
    \end{subfigure}%
    \begin{subfigure}{.33\textwidth}
    \centering
    \includegraphics[width=0.78\textwidth]{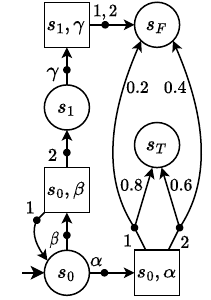}
    \caption{}
    \label{fig:example:game-abstraction}
    \end{subfigure}%

\caption{
(a)~Family $\family$ of two MDPs.
(b)~Quotient MDP $\qmdp_\family$ for the family $\family$.
(c)~Game abstraction for family $\family$. Circles (squares) denote states of Player 1 (2). Transitions without explicit probability have a probability of 1. States $s_T$ and $s_F$ are absorbing, and their action with self-loop is omitted.
}
\label{fig:example}
\end{figure}

\vspace{-1em}
\subsection{Problem Statement on Compact Representations}

We aim to compute a policy map for a family of MDPs, assuming compact input and output representations, which we formalise below.

\subsubsection{Compact input} An explicit, tabular representation of a family of MDPs is unsuitable to prevent algorithms from scaling linearly with the number of MDPs. Furthermore, such representations do not carry the necessary structural information to implement efficient heuristics. 
As we show later in Sec.~\ref{sec:experiments}, a family of MDPs can be represented using a high-level formalism like the PRISM language~\cite{KNP11}.
However, for the development of this paper, we instead consider an operational model. We use a \emph{quotient MDP}, which is a mild variation of feature MDPs~\cite{DBLP:journals/fac/ChrszonDKB18} and an extension of quotient MDPs for families of Markov chains~\cite{cegar}.

\begin{definition}[Quotient MDP]
\label{def:quotient}
Let $\family= \indexedmdps[\in \indices]$ be a family of MDPs.
The \emph{quotient MDP} is a pair $(\qmdp_\family,\coloring)$, where $\qmdp_\family = (\states,\sinit, Act_\qmdp,\mpm_\qmdp)$ is an MDP over the set $Act_\qmdp = \actions \times 2^{\indices}$ of actions where $Act_\qmdp(s) = \{ (\act,\subindices)\ \mid \act \in Act(s), \ \subindices \in \indicespartition \}$; action $(\act,\subindices)$ will be denoted as $\act_\subindices$; action $\act_{\{i\}}$ will be denoted as $\act_i$ for brevity.
The transition function is $\mpm_\qmdp(s,\act_I,s') = \mpm_{i} (s,\act,s')$ where $i \in \subindices$.
Function $\coloring\colon \act_I \mapsto I$ maps action to the identifiers.
\end{definition}
Essentially, quotient MDP $\qmdp_\family$ is an MDP that can execute in state $s$ action $\act$ from an arbitrary $M_i \in \family$.
We omit the subscript from $\qmdp_\family$ whenever family \family is clear from the context. The shape $(\act,\subindices)$ of actions allows us to efficiently encode families of MDPs where action $\act$ coincides in many family members. Fig.~\ref{fig:example:quotient} illustrates this, where we merge action $\gamma$ shared by MDPs $1,2$ into $\gamma_{1,2}$.

We say that the policy $\sched \in \Sigma^\qmdp$ is \emph{consistent (in identifier $i$)} iff $\exists i \in \indices \, \forall s \in \states \colon i \in \coloring(\sched(s))$, i.e. if the policy only selects actions from some MDP $M_i$.

\begin{lemma}\label{lemma:quotient-maxmax}
    Consider family $\family= \indexedmdps[\in \indices]$ of MDPs and its quotient MDP $\qmdp_\family$.  
    $$\probmax[\qmdp_\family]{\F T} < \lambda \quad\text{implies}\quad \forall M_i \, \forall \sigma \in \Sigma^M \ \colon \reach{M_i^\sigma} < \lambda$$
    and thus for the threshold $\geq \lambda$ and all $M_i$, it holds that ${\pmp(M_i) = \unsat}$\footnote{Similarly, 
    $\probmin[\qmdp_\family]{\F T} \geq \lambda$ implies $\forall M_i \, \forall \sigma \in \Sigma^M \ \colon \reach{M_i^\sigma} \geq \lambda.$}. 
    
\end{lemma}

\paragraph{Proof.} We provide proof of Lemma~\ref{lemma:quotient-maxmax} in Appendix~\ref{app:proofs}.

\vspace{-0.2em}
\begin{example}\label{ex:quotient}
    Consider family $\family$ of MDPs with quotient MDP $\qmdp_\family$ from Fig.~\ref{fig:example:mdp-family} and \ref{fig:example:quotient}, respectively, and consider specification $\probthreshold[M]{\geq \lambda}{\F s_T}$. Model checking $\qmdp_\family$ yields $\probmax[\qmdp_\family]{\F s_T} = 0.8$ under policy $\sigma(s_0) = \act_1$. Thus, if $\lambda = 0.9$, we know that all MDPs in $\family$ are unsatisfiable. However, if $\lambda = 0.5$, from $\mathbb{P}_{\max}{=}0.8$ we cannot deduce anything, even though all MDPs in $\family$ are satisfiable.
\end{example}

\subsubsection{Compact output} We aim to pair a compact representation of the family with a compact representation of its policy map. In this paper, we use \emph{policy trees}\footnote{To simplify the exposition of this paper, we define binary trees. In our implementation, we use their natural extension to $n$-ary trees.}.
\begin{definition}[Policy tree]
Let $\family = \indexedmdps[\in \indices]$ be a family of MDPs and $\varphi$ be a specification. The (satisficing) \emph{policy tree} is a tuple $\ptree = (V,l,r,\nodelabel,\leaflabel)$ where $(V,l,r,)$ is a binary tree over the set $V$ of vertices and $l,r \colon V \nrightarrow V$ are functions for the left and right successors of each inner node, respectively.
$\nodelabel \colon V \rightarrow 2^\indices$ associates nodes with subfamilies, and  $\leaflabel \colon V \nrightarrow \fschedulers \ {\cup} \ \{\unsat\}$  associates leaf nodes with policies. The root $R \in V$ of the tree is associated with the complete family: $\nodelabel(R) = \indices$. Furthermore, for any node $n \in V$, either:
\begin{enumerate}
    \item $n$ is an inner node, and  $\nodelabel(l(n))$ and $ \nodelabel(r(n))$ are a disjoint partition of~$\nodelabel(n)$.
    \item $n$ is a leaf with $\leaflabel(n) \in \Sigma^M$. Then $\leaflabel(n)$ is a robust policy for $\{M_i\}_{i \in \nodelabel(n)}$.
    \item $n$ is a leaf with $\leaflabel(n) = \unsat$. Then $\pmp(M_i) = \unsat$ for all $M_i \in \{M_i\}_{i \in \nodelabel(n)}$.
\end{enumerate}
\end{definition}
These definitions allow us to state the central problem of this paper.

\begin{mdframed}
\textbf{Policy tree synthesis}: Given a family of MDPs $\family= \indexedmdps[\in \indices]$ represented as a quotient MDP  \qmdpT, and a specification $\varphi$, construct the policy tree~$\ptree$.
\end{mdframed}

\subsubsection{Existence of small policy trees} Our goal is to construct for a given family $\family$ a small policy tree $\ptree$. There are two implicit assumptions that admit a small~$\ptree$: 1) The particular MDPs $M_i$ describe similar scenarios (recall Fig.1) and share the state and action space. It is thus natural that a good policy for one MDP performs (reasonably) well on another similar MDP. 2) $\ptree$ contains satisficing policies for the given threshold (on the probability/reward) rather than optimal policies for every MDP. The lower thresholds typically impose weaker constraints, while thresholds close to one may cause the specification to be unsatisfiable for a majority of MDPs -- both cases admit small policy trees. 

\section{Game Abstraction for MDP Families}
\label{sec:AR}

Let us first motivate the game abstraction by demonstrating the key characteristic of the quotient MDP $\qmdp_\family$: Lemma~\ref{lemma:quotient-maxmax} and Example~\ref{ex:quotient} showcase that model checking $\qmdp_\family$ allows reasoning about non-alternating specifications. However, the result $\reachmax{\qmdp_\family} \geq \lambda$ does not give sufficient information about the entire $\family$: it may still hold that some or even all $M_i\in \family$ do not meet $\geq \lambda$. This is because quotient $\qmdp_\family$ lumps both sources of nondeterminism~--~action and identifier selection~--~together. To remedy this, we leverage an abstraction based on stochastic games, as shown in the following example:


\begin{example}\label{ex:game}
    Following Example~\ref{ex:quotient} for $\lambda = 0.5$, we can obtain more precise information by using the game abstraction $\game_\family$ (see Fig.~\ref{fig:example:game-abstraction}). Solving $\game_\family$ yields value $\playeronemax[\game_\family]{\F{s_T}} = 0.6 \geq \lambda$ and therefore we know all MDPs in $\family$ are satisfiable. In fact, from Theorem~\ref{theorem:policy} (below)
    we know that policy $\sched[1]$ for Player 1 with $\sched[1](\sinit)=\act$ is a robust policy for $\family$.
\end{example}

\subsection{Background on Stochastic Games}

We recap some background on stochastic games. See e.g.~\cite{SK16,doi:10.1073/pnas.39.10.1095} for more details.

\begin{definition}[Stochastic game] A \emph{stochastic game} (SG) is a tuple $G=(M,S_1,S_2)$ where $M= \mdpT$ is the \emph{underlying MDP} and $(S_1,S_2)$ is a partition of the set~$S$ of states into Player~1 states and Player~2 states, respectively.     
\end{definition}

The state partition naturally partitions polices $\schedulers$ (in our case, memoryless deterministic policies) into $\schedulers_1$ and $\schedulers_2$ where $\sigma_1 \in \schedulers_1$ maps $S_1$ to $\Act$ and $\sigma_2 \in \schedulers_2$ maps $S_2$ to $\Act$.
A pair of policies $\sigma_1 \in \schedulers_1$ and $\sigma_2 \in \schedulers_2$ describes a policy on the underlying MDP. We denote the MC induced by such a pair with $G^{\sigma_1,\sigma_2}$. In what follows, we assume that Player 1 \emph{maximises} the reachability probability and Player 2 \emph{minimises} the probability.

The \emph{Player 1 maximum reachability probability} of the set~$T$ of target states from the initial state $\sinit$ in $G$ is defined as\footnote{The minimal reachability probability for Player 1 is defined analogously.}
\[\textstyle \playeronemax[G]{\F{T}} \coloneqq \sup_{\sigma_1 \in \schedulers_1} \inf_{\sigma_2 \in \schedulers_2} \reach{G^{\sigma_1,\sigma_2}}.\] 
SGs with reachability objectives are determined and thus:
\[\textstyle
\sup_{\sigma_1 \in \schedulers_1} \inf_{\sigma_2 \in \schedulers_2} \reach{G^{\sigma_1,\sigma_2}} = \inf_{\sigma_2 \in \schedulers_2} \sup_{\sigma_1 \in \schedulers_1} \reach{G^{\sigma_1,\sigma_2}}.\]
It has also been shown that for these objectives, optimal policies for both players exist, and memoryless deterministic
policies suffice. In particular, the probability $\playeronemax[G]{\F{T}}$ is achieved by policy $\sigma = (\sched[1], \sched[2])$ where $\sched[1]$ is the policy of Player 1 and $\sched[2]$ is the policy of Player~2. No polynomial-time algorithm is known for reachability objectives. However, the worst-case exponential-time algorithms based on value/policy iteration work reasonably well on many practical problems~\cite{KRETINSKY2022104885} and are available in tools like Prism-Games~\cite{KNPS20} and Storm~\cite{STORM}. 

\subsection{Quotient Game Abstraction}
\label{subsec:game}
We now define the SG abstraction of $\family$, allowing us to effectively reason about the winning policies in $\family$. Without loss of generality, assume a threshold $\geq \lambda$. Intuitively, the game abstraction of $\family$ is a two-player turn-based stochastic game where Player~1 in state $s$ chooses one of the actions $\act \in \Act(s)$ and the Player~2 subsequently chooses one of the MDPs $M_i \in \family$ in which $\act$ will be~executed. 

\begin{definition}[Game abstraction]
\label{def:game}
Let $\family= \indexedmdps[\in \indices]$ be a family of MDPs. The \emph{game abstraction} for $\family$ is a stochastic game $\game_\family = (M, S_1, S_2)$ with $S_1 = \states$, $S_2 = \states {\times} \actions$ and an underlying MDP $M = (S_1 \cup S_2,\sinit, \Act_{\game},\mpm_{\game})$ where
$\Act_{\game} = \actions \cup 2^\indices$.
For each $s \in S_1$,  $\Act_{\game}(s) = \actions(s)$ and $\forall \act \in \Act(s) \colon \mpm_{\game}(s, \act, (s, \act)) = 1$;
For each $(s,\act) \in S_2$, $\Act_{\game}(s,\act) = \indicespartition$ and $\forall \subindices \in \indicespartition: \mpm_{\game}((s, \act), \subindices) = \mpm_{i}(s, \act)$ where $i \in \subindices$.
\end{definition}
We omit the subscript from $\game_\family$ whenever family \family is clear from the context.
We say that Player 2 policy $\sched[2] \in \schedulers_2$ is \emph{consistent (in identifier $i$)} iff $\exists i \in \indices \, \forall s \in S_2 \colon i \in \sched[2](s)$, i.e.~if Player 2 only selects actions from some MDP $M_i$.
We say that the game policy is consistent iff its Player 2 policy is consistent.

\begin{lemma}
\label{lemma:player2-consistency}
If 
$\sched[2]$ is consistent in $i$ then
$\forall \sched[1] \colon \reach{\game_{\family}^{\sched[1],\sched[2]}} = \reach{M_i^{\sched[1]}}$\footnote{The MDP $\game^{\sched[2]}$ is actually weakly bisimilar to $M_i$.}.
Consequently, assuming $\playeronemax[\game_\family]{\F{T}}$ is achieved using $\sched[1]$ and $\sched[2]$: \[ \text{$\sched[2]$ is consistent in $i$}\quad\text{implies}\quad\text{$\reachmax{M_i} = \reach{M_i^{\sched[1]}}$}.\]
\end{lemma}

Lemma~\ref{lemma:player2-consistency} shows the connection between the game abstraction $\game$ and the MDPs in family $\family$. We use this lemma as a basis for the following theorem:

\begin{theorem}[Policy from winning game]\label{theorem:policy}
Consider a family $\family= \indexedmdps[\in \indices]$ and its game abstraction $\game_\family$. Assume $\playeronemax[\game_\family]{\F{T}}$ is achieved by policy $\sched[1]$. 
\[ \playeronemax[\game_\family]{\F{T}}  \geq \lambda \quad\text{implies}\quad \forall M_{i} \in \family : \reach{M_i^{\sched[1]}} \geq \lambda.\]
\end{theorem}
\paragraph{Proof.} We provide proof of Theorem~\ref{theorem:policy} in Appendix~\ref{app:proofs}.


\paragraph{The game abstraction is a proper abstraction} and thus $\playeronemax[\game]{\F{T}}  < \lambda $ does not imply that an unsatisfiable MDP exists in the family, nor does it imply that the robust policy does not exist for \family. The reason is that Player 2 is too powerful as they are not bound to play consistently and can adapt their selection of an MDP based on the state-action selection of Player 1. In our illustrative example, this means that the chair can be placed based on the selected route of the agent.

However, if $\playeronemax[\game]{\F{T}}  < \lambda$ under the optimal policies $\sched[1]$ and $\sched[2]$ where $\sched[2]$ is consistent in identifier $i$, then from Lemma~\ref{lemma:player2-consistency} we know that $\reachmax{M_i} < \lambda$. We use this later in our splitting strategy.

Theorem~\ref{theorem:policy} also demonstrates that the game abstraction can only find robust policies. As we show in the next section, solving the game abstraction is an adequate sub-routine for the construction of the satisficing policy tree.

\begin{remark}
Consider a relaxed version of our problem statement that aims to synthesise a Boolean map $\bmp\colon \family \rightarrow \{\top,\unsat\}$ such that $\bmp(M_i) = \top$ iff there is a winning policy for $M_i$. 
Clearly, the game abstraction, yielding robust policies, is a very coarse abstraction. The existence of more refined abstractions tailored to this relaxed problem is outside the scope of this paper. 
\end{remark}

\begin{remark}
\label{rem:ordering}
We designed the game abstraction such that Player 1, choosing actions, chooses before Player 2, choosing identifiers, which leads to the use of auxiliary Player~2 states consisting of state-action pairs. 
This is the action-identifier (AI) ordering. There naturally also exits the identifier-action (IA) ordering where Player 2 chooses identifiers $\subindices$ before Player~1 selects an action. Consequently, those games have auxiliary states  $(s,\subindices)$. 
In general, the maximisation in the IA ordering can lead to a higher (i.e. better) probability than the AI ordering -- the maximiser (Player~1) can have a better reaction after the Player~2 action selection. However, the IA ordering requires, for each state $s\in S$, up to $|\family|$ auxiliary states, which is typically significantly more compared to the AI ordering that requires only up to $Act(s)$ states. Thus, we use the AI ordering\footnote{Game-based abstraction for MDPs~\cite{DBLP:conf/tacas/HahnHWZ10,DBLP:journals/fmsd/KattenbeltKNP10} prefer the IA ordering as it is tighter.}.
\end{remark}

\section{Recursive Abstraction-Refinement for Policy Trees}
\label{sec:algorithm}

We first propose a divide-and-conquer algorithm for the policy tree synthesis that builds on the game abstraction and briefly discuss alternative sub-routines for the synthesis algorithm. Further, we propose several splitting strategies that drive the abstraction refinement. Finally, we introduce an essential post-processing routine that reduces the size of the police tree.

\vspace{-0.5em}
\subsection{Divide-and-Conquer Algorithm}
Our algorithm creates a (binary) tree, where leaves contain policies and inner nodes point to two sub-trees, annotated with the subfamily for which they describe the winning policies. For an open node describing family $\family=\indexedmdps[\in \indices]$, we first aim to find a robust policy for \family (\textsc{findRobust} call on l.~\ref{alg:call:findRobust}). If we succeed, we can return a leaf node annotated with the policy.
If we do not succeed, we try to prove that there are no winning policies for any of the MDPs (\textsc{testUnsat} call on l.~\ref{alg:call:testUnsat}), in which case we return a leaf node annotated with \unsat. 
In all other cases (l.~\ref{alg:call:CreateSplit}), we split our family into two subfamilies, guided by feedback from \textsc{findRobust} or \textsc{testUnsat} subroutines, as discussed later. We then create an inner node that points to two recursively constructed trees. 

Implementations of \textsc{createSplit} are discussed in Sec.~\ref{sec:algorithm:splitting}.
The default variant of \textsc{findRobust} sub-routine (ll.~\ref{alg:findRobust:start}-\ref{alg:findRobust:end}) is based on the game abstraction, where we solve the game abstraction $\game$ for family \family and return the optimising policy $\gsched$
If $\playeronemax[\game]{\F{T}} \geq \lambda$, then $\gsched$ is robust for \family by Theorem~\ref{theorem:policy}; Otherwise, we use $\gsched$ for splitting.
In \textsc{testUnsat} (ll.~\ref{alg:testUnsat:start}-\ref{alg:testUnsat:end}), we leverage the quotient~$\qmdp$: we check whether $\reachmax{\qmdp} < \lambda$ ensuring that $\forall M_i \, \forall \sigma \in \Sigma^M: \reach{M_i^\sigma} < \lambda$.


We remark that a fresh construction of the quotient from Def.~\ref{def:quotient} and the game from Def.~\ref{def:game} for the subfamily $\family' \subset \family$ can be avoided as the actions in the quotient/game constructed for $\family'$ are a subset of the actions in the quotient/game for \family. Thus, instead of rebuilding the quotient/game, we only restrict the allowed actions.

\begin{algorithm}[t]
\DontPrintSemicolon
\SetKwFunction{findRobust}{\textsc{findRobust}}
\SetKwFunction{testUnsat}{\textsc{testUnsat}}
\SetKwFunction{buildTree}{\textsc{buildTree}}
\SetKwInOut{Input}{Input}
\SetKwInOut{Output}{Output}
\SetKw{Continue}{continue}
\SetKw{Yield}{yield}
\SetKwComment{Comment}{$\triangleright$\ }{}
\SetKwData{Data}{Global}
\Data{family $\indexedmdps[\in \indices]$ of MDPs, target set $T$, threshold $ \geq \lambda$}\\
\Input{Index set $\indices$}
\Output{Root of the tree representation $\ptree$ of the policy map $\pmp$}
\BlankLine
\SetKwProg{myfunction}{Function}{}{}
\myfunction{\textsc{buildTree}{$(\indices)$}}{
$sat, \gsched \gets$~ \textsc{findRobust}{$(\indexedmdps[\in \indices])$} \; \label{alg:call:findRobust}
\lIf{sat}{\Return{ $\textsc{LeafNode}(\indices, \gsched)$ }}
 \lIf{\textsc{testUnsat}$(\indexedmdps[\in \indices])$}{\Return{$\textsc{LeafNode}(\indices, \unsat)$}} \label{alg:call:testUnsat}
 $\indices' \gets \textsc{CreateSplit}(\indices,\gsched)$ \; \label{alg:call:CreateSplit}
 \Return {$\textsc{InnerNode}(\indices$, \textsc{buildTree}{$(\indices')$}, \textsc{buildTree}{$(\indices {\setminus} \indices'$} $)$}
}
\BlankLine
\myfunction{\textsc{findRobust}{$(\family)$}} {
\label{alg:findRobust:start}
$\game \gets \textsc{buildGame}(\family)$ \Comment{Applying Def.~\ref{def:game}} 
$res,\gsched  \gets \textsc{compute}(\playeronemax[\game]{\F{T}})$ \Comment{SG model-checking  }
        \Return{$res \geq \lambda, \gsched$} \Comment{$\gsched$ is the optimal policy}
\label{alg:findRobust:end}
}

\myfunction{\textsc{testUnsat}{$(\family)$}}{
\label{alg:testUnsat:start}
$\qmdp \gets \textsc{buildQuotient}(\family)$ \Comment{Applying Def.~\ref{def:quotient}} 
$res, \qsched \gets \textsc{compute}(\reachmax{\qmdp})$ \Comment{MDP model-checking  }
        \Return{$res < \lambda, \qsched$} 
\label{alg:testUnsat:end}
}

\caption{Recursive tree construction}
\label{alg:general}
\end{algorithm}


\begin{theorem}\label{theorem:correctness}
Assuming the subroutines terminate and that \textsc{CreateSplit} returns non-trivial subsets,
Algorithm~\ref{alg:general} is sound and complete and needs at most $n$ splits, where $n$ is the number of MDPs in the family.
\end{theorem}
\vspace{-0.5em}
\paragraph{Proof.} We provide proof of Theorem~\ref{theorem:correctness} in Appendix~\ref{app:proofs}.

\subsection{Alternative Strategies for Finding Robust Policies}
\label{sec:algorithm:randomization}

As already discussed in Sec.~\ref{subsec:game}, in our game abstraction, Player 2 is very powerful since they can adapt their selection of an MDP during the game. In the experimental section, we show that in some models this hinders the performance of the synthesis algorithm. We also consider a different abstraction where Player 2 is replaced by the randomised selection of identifiers $\subindices$, i.e. by the policy $\schedrandom$ that in every state $(s,\act) \in S_2$ uniformly selects one of the identifier classes $\subindices \in \indicespartition$. In the game $\game$, $\schedrandom$ thus induces an MDP $\game^{\schedrandom}$. The maximising policy $\sigma_{\max}$ for $\game^{\sigma_{rnd}}$ indeed does not provide any guarantees on the existence or non-existence of the robust policy. Thus we need also a $\textsc{testSat}(\family, \sigma_{\max}$) sub-routine that checks whether $\sigma_{\max}$ is robust. On the other hand, model checking $\game^{\sigma_{rnd}}$ is computationally less demanding than solving the game $\game$, and the randomisation ensures that $\sigma_{\max}$ is aware of all possible MDPs~$M_i$.


Implementing $\textsc{testSat}$ using an enumeration of all $M_i \in \family$ is not tractable. We instead use a more sophisticated approach leveraging the recent work on the inductive synthesis of MCs~\cite{andriushchenko2021inductive}. The candidate policy $\sigma_{\max}$ induces in the family \family a family of MCs. The inductive synthesis methods based on abstraction refinement and counter-example generalisation enable an efficient analysis of this family of MCs. In particular, we can find MCs satisfying the property that corresponds to the MDPs where $\sigma_{\max}$ is winning.
Our preliminary experiments suggested that running MC synthesis loop within each iteration is too expensive.
Instead, we run MC synthesis in a lightweight but incomplete fashion: $\sigma_{\max}$ is accepted only if $\reach{\qmdp^{\sigma_{\max}}} \geq \lambda$.


\subsection{Splitting strategies}
\label{sec:algorithm:splitting}

We design two splitting strategies. \emph{Pessimistic} splitting is an adoption of an approach for families of MCs~\cite{cegar} and is particularly suitable for splitting families containing both satisfiable and unsatisfiable MDPs. \emph{Optimistic} splitting is geared towards increasing the likelihood that we find a robust policy.
To present both approaches, we will need the following definitions.
Given MDP~$M$ and a policy $\sched \in \schedulers$, the \emph{reachable fragment (under~$\sched$)} is a set $\reachable(M,\sched)$ of states reachable in $M^{\sched}$ from the initial state $\sinit$. Given quotient MDP $\qmdpT$, a policy $\qsched \in \Sigma^{\qmdp}$ is \emph{consistent in identifier $i$ on the reachable fragment} iff $\forall s \in \reachable(\qmdp,\qsched) \colon i \in \coloring(\qsched(s))$. Let $\consdindices(\qsched) \subseteq \indices$ be the set of all identifiers in which $\qsched$ is consistent on the reachable fragment.
To \emph{split} the input index set $\indices$ is to partition it into $\indicesone$ and $\indicestwo \coloneqq \indices {\setminus} \indicesone$; let $\familyone,\familytwo$ denote the corresponding partition of the family $\family = \indexedmdps[\in \indices]$ of MDPs.

\paragraph{Pessimistic splitting.}
The goal of the pessimistic splitting is to partition the family $\family$ into $\familyone,\familytwo$ s.t.~at least one of the corresponding quotient MDPs $\qmdp_{\familyone},\qmdp_{\familytwo}$ is a strict restriction of $\qmdp_{\family}$.
To achieve this, we inspect the policy $\qsched$, computed in $\textsc{testUnsat}()$ for the quotient MDP $\qmdp_\family$, and compute the set $\consdindices(\qsched)$. 
If $\consdindices(\qsched) {\neq} \emptyset$, we set $\indicesone {=} \consdindices(\qsched)$. This will guarantee that policy $\qsched$ is \emph{robust} for $\familyone$\footnote{
$\qmdp_{\family}^{\qsched}$ is the same MC as $M_i^{\overline{\qsched}}$, where $\overline{\qsched}(s) = \act$ iff $\qsched(s) = (\act,\cdot)$, for any $i \in \consdindices(\qsched)$. Thus, if $\reach{\qmdp_{\family}^{\qsched}}\geq \lambda$, then $\overline{\qsched}$ must be winning for $M_i$.
} and that $\qsched \notin \Sigma^{\qmdp_{\familytwo}}$. 
Otherwise, if $\consdindices(\qsched) {=} \emptyset$, i.e.~policy $\qsched$ is not consistent, there must exist identifiers $i,j \in \indices, i \neq j$ s.t.~$\exists s_i,s_j \in \reachable(\qmdp_{\family},\qsched) \colon i \notin \coloring(\qsched(s_j)) \textrm{ and } j \notin \coloring(\qsched(s_i))$. Then, in most cases, the splitting where $i \in \indicesone$ and $j \in \indicestwo$ leads to sub-quotients $\qmdp_{\familyone},\qmdp_{\familytwo}$ that are both strict restrictions of $\qmdp_{\family}$.
The name \emph{pessimistic} alludes to the fact that this approach tries to separate family \family into subfamilies $\familyone$ and $\familytwo$ that contain only satisfiable and only unsatisfiable MDPs, respectively.
When we use the $\game^{\schedrandom}$ from Section~\ref{sec:algorithm:randomization}, we use pessimistic splitting.

\paragraph{Optimistic splitting.} Optimistic splitting executes the same reasoning as above, but, instead of MDP policy $\qsched$, uses game policy $\gsched$ computed in \textsc{findRobust}.
The notion of the reachable fragment $\reachable(\game,\gsched)$ in the game \game is defined analogously. We say that game policy $\gsched = (\sched[1],\sched[2])$ is \emph{consistent in identifier $i$ on the reachable fragment} iff $\forall s \in \reachable(\game,\gsched) \cap S_2 \colon i \in \sched[2](s)$; recall that $S_2$ denotes the states of Player~2.
Similarly, let $\consdindices(\gsched)$ denote the set of all identifiers in which $\gsched$ is consistent on the reachable fragment.
If $\consdindices(\gsched) \neq \emptyset$, then we set  $\indicesone = \consdindices(\gsched)$; Note that in this case, this \emph{does not} guarantee that family $\familytwo$ contains only unsatisfiable MDPs (see Example~\ref{ex:unreachable} below), although it still holds that $\gsched \notin \Sigma^{\game_{\familytwo}}$.
Otherwise, if $\gsched$ is inconsistent on the reachable fragment, we split using distinguishing identifiers $i,j$ as above.



\begin{figure}[t]
    \centering
    \begin{subfigure}[t]{0.35\textwidth}
        \centering
        \includegraphics[width=\textwidth]{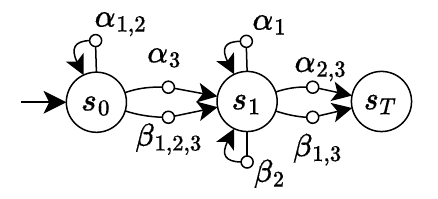}
        \label{fig:unreachable:quotient}
    \end{subfigure}
    \begin{subfigure}[t]{0.60\textwidth}
        \centering
        \includegraphics[width=\textwidth]{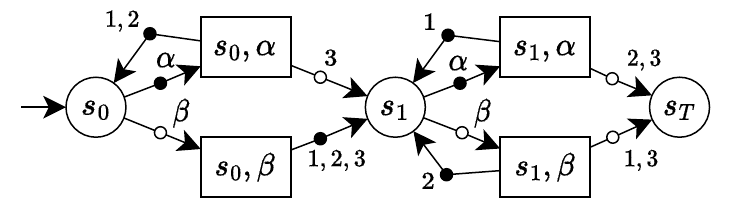}
        \label{fig:unreachable:game}
    \end{subfigure}
    \vspace{-1.5em}
    \caption{Quotient (left) and game abstraction (right) for a simple family of MDPs.}
    \label{fig:unreachable}
    \vspace{-1.5em}
\end{figure}

Surprisingly, the optimistic splitting can profit from taking into account states that are unreachable under $\gsched$, as illustrated in the following example.



\begin{example}
\label{ex:unreachable}
Fig.~\ref{fig:unreachable} (left) depicts a quotient for a family of three MDPs. The goal is to reach $s_T$ with probability 1. All three MDPs are satisfiable: policy $\sigma(s_0){=}\sigma(s_1){=}\beta$ is winning for $M_1$, policy $\sigma'(s_0){=}\beta,\sigma'(s_1)=\alpha$ is winning for $M_2$, and any policy is winning for $M_3$.
There is no robust (memoryless) policy. 
Thus, the smallest policy tree contains leaf nodes labelled with policies $\sigma,\sigma'$.

Fig.~\ref{fig:unreachable} (right) depicts the corresponding game abstraction. The optimal value in this game is $0$ and a corresponding policy is visualised via filled action circles.
$\gsched$ is consistent in identifiers 1,2 on the reachable fragment, however, splitting into $\{1,2\}$ and $\{3\}$ is bad as there is no robust policy for $\{M_1,M_2\}$. Instead, we inspect $\gsched$ on all states and see that it is not consistent due to $\sigma_2(s_1,\alpha)=1$ and $\sigma_2(s_1,\beta)=2$. Separating~1 and~2 into different subfamilies (e.g. $\{1,3\} \cup \{2\}$) allows the game abstraction to yield robust policies for both subfamilies.
\end{example}






\subsection{Post-Processing the Policy Tree}
\label{sec:post-processing}

We propose a post-processing step that reduces the size of the policy tree based on similarities between policies generated for particular leaf nodes. We say that policies $\sched[i],\sched[j]$ associated with families $\family_i,\family_j$ are \emph{compatible} if they agree on their choices in states that are in both reachable fragments: $\forall {s \in S} \colon$ $ \left( s \in \reachable(\qmdp_{\family_i},\sched[i]) \right)$ $ \land \left( s \in \reachable(\qmdp_{\family_j},\sched[j]) \right) \Rightarrow \sched[i](s){=}\sched[j](s)$.
We also introduce the following policy merging operation: $\sched[i+j](s) = \sched[i](s)$ if  $s \in \reachable(\qmdp_{\family_i},\sched[i])$, and $\sched[i+j](s) = \sched[j](s)$ otherwise.
%
%
%
Note that if policies $\sched[i]$ and $\sched[j]$ are compatible, then $\sched[i+j]$ and $\sched[j+i]$ differ only in states that are unreachable in both $\qmdp_{\family_i}^{\sched[i]}$ and $\qmdp_{\family_j}^{\sched[j]}$.
The proposed post-processing algorithm executes the following three steps.

\paragraph{Step 1.} For each pair of sibling nodes associated with families $\family_l,\family_r$ and robust policies $\sched[l],\sched[r]$, we attempt to verify whether $\sched[l]$ is a suitable policy for $\family_r$ by constructing $\sched[l+r]$ and checking whether $\probmin[\qmdp_{\family_r}^{\sched[l+r]}]{\F T} \geq \lambda$, i.e.~if $\sched[l+r]$ is robust for $\family_r$ (see Lemma~\ref{lemma:quotient-maxmax}).
If this policy is robust for $\family_r$, we label both nodes with policy $\sched[l+r]$. Otherwise, we try to verify whether $\sched[r+l]$ works for $\family_l$.

\paragraph{Step 2.}
We inspect each pair $n_i,n_j$ of leaf nodes. If they are labelled with compatible policies $\sched[i],\sched[j]$, then 
the policy $\sched[i+j]$ must be winning for both $\family_i$~and~$\family_j$ and we label both nodes with $\sched[i+j]$.


\paragraph{Step 3.} If for any inner node $n$ it holds that $\leaflabel(l(n)) = \leaflabel(r(n))$, that is, both its children are labelled as unsatisfiable or both labelled with the same policy, we remove nodes $l(n),r(n)$ and set $\leaflabel(n) = \leaflabel(l(n))$.

\begin{remark}
   A final step would be to convert the resulting policy tree to a directed acyclic graph by merging \emph{all} nodes labelled with the same policy and \emph{all} nodes associated with $\unsat$. In our experimental evaluation, we avoid this step because the number of leaves in the resulting policy tree gives us a useful measure of how the size of the tree relates to the size of the family of~MDPs. 
\end{remark}

\begin{figure}[t]
    \centering
    \begin{subfigure}[t]{0.45\textwidth}
        \centering
        \includegraphics[width=0.9\textwidth]{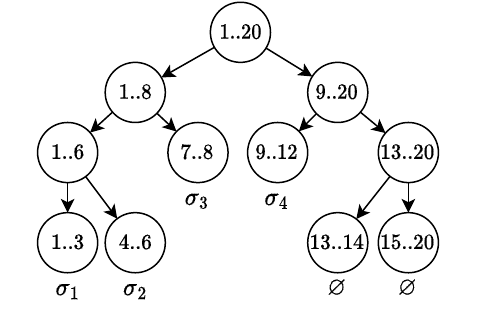}
    \end{subfigure}\hspace{2em}
    \begin{subfigure}[t]{0.35\textwidth}
        \centering
        \includegraphics[width=\textwidth]{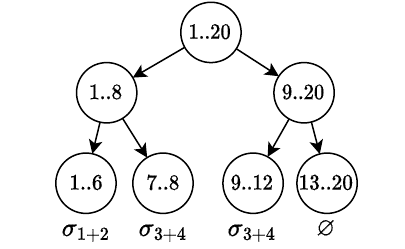}
    \end{subfigure}
    \caption{Policy tree before (left) and after (right) post-processing.}
    \label{fig:post-processing}
    \vspace{-1em}
\end{figure}

 \vspace{-1em}
\begin{example}
Policy tree post-processing is illustrated in Fig.~\ref{fig:post-processing}. In Step 1, we find out that policy $\sched[1+2]$ is robust for family $\{4..6\}$, thus allowing us to label the whole family $\{1..6\}$ with $\sched[1+2]$.
In Step 2, we merge compatible policies $\sigma_3$ and $\sigma_4$ into $\sched[3+4]$.
Lastly, we merge nodes corresponding to families $\{13..14\}$ and $\{15..20\}$ as they are siblings that are both unsatisfiable. In this example, we reduced the policy tree from $11$ nodes, $6$ leaves and $4$ policies to $7$ nodes, $4$ leaves and $2$ policies.
A larger example of a policy tree obtained for one of the benchmarks from the experimental evaluation can be found in Appendix~\ref{app:tree}.
\end{example}

\vspace{-1em}
\section{Experimental Evaluation}

\label{sec:experiments}
We implement the proposed algorithm and its variants on top of \paynt~\cite{DBLP:conf/cav/AndriushchenkoC21} and \storm~\cite{STORM}. The implementation and all the considered benchmarks are publicly available\footnote{\url{https://doi.org/10.5281/zenodo.12569976}}.
Our evaluation focuses on the following two questions:

\begin{enumerate}
\item[Q1:] \emph{Can we generate policy trees for large families? Do we find winning policies for families that are beyond the reach of the baselines?}
\item[Q2:] \emph{Can we find policy trees with significantly fewer leaves than the family size?} 
\end{enumerate}
Evaluation is supplemented with an ablation study considering the alternative strategies for finding robust policies (Sec.~\ref{sec:algorithm:randomization}) and splitting (Sec.~\ref{sec:algorithm:splitting}). 

\subsubsection{Setup}
We describe benchmarks and algorithm selection below. Each experiment is run on a single core of an Intel i5-12600KF @4.9GHz CPU with 64GB of RAM. We limit memory to 16GB and time to 2 hours.

\paragraph{Benchmarks.}
 We evaluated the methods on various MDP sketches based on a wide range of MDPs from the literature~\cite{dpm-1,uav,rover,grids}. We obtained these sketches by introducing ``holes" into the MDP Prism programs (see Appendix~\ref{app:sketching}). We believe these families of MDPs describe interesting (high-level) planning and decision-making problems. The benchmarks were selected to showcase the main advantages and drawbacks of the proposed approach.
 The left part of Tab.~\ref{tab:main} reports the main statistics of the models, including the number $|\family|$ of family members and the percentage of satisfiable MDPs. We also report the number of states $|S_\family|$ and the total number $\sum Act \coloneqq \sum_{s}Act^{\qmdp}(s)$ of actions in the quotient MDP $\qmdp$. In our benchmarks, $|S_\family|$ is close to the reachable fragment in every $M_i \in \family$ except for the $\emph{dodge-8}$ models where the average size of the MDPs is ${\sim}$20-fold smaller than $|S_\family|$. 



\paragraph{Algorithms.}
We consider two baseline algorithms:
i) Enumerating the MDPs \emph{one-by-one} and solving each MDP individually. This approach produces a list of solutions, explicitly holding results for each MDP. ii) Creating \emph{all-in-one} MDP that is stored symbolically using decision diagrams and solving it. For both approaches, we use Storm~\cite{STORM}\footnote{The symbolic engines of Storm or PRISM do not actually export policies.}.
In all algorithms, we use a policy iteration for solving SGs and value iteration for MDPs. Our algorithm exports a policy tree that compactly represents the set of winning policies.


\begin{table}[t]
\renewcommand{\tabcolsep}{1.5pt}
\scalebox{0.97}{
\begin{tabular}{l cccr@{\hskip 12pt}SSSr@{\hskip 12pt}rr}
\toprule

\multirow{2}{*}{model} &
\multicolumn{4}{c@{\hskip 12pt}}{model info} &
\multicolumn{4}{c@{\hskip 12pt}}{our approach} &
\multicolumn{2}{c}{speedup wrt.}\\
\cmidrule(lr{1.75em}){2-5}\cmidrule(lr{1.75em}){6-9}\cmidrule(lr){10-11}

&
$|S_\family|$ &
$\sum Act$ &
$|\family|$ &
\multicolumn{1}{c@{\hskip 12pt}}{SAT\,\%} &
\multicolumn{1}{c}{L/$\family$\,\%} &
\multicolumn{1}{c}{P/SAT\,\%} & 
\multicolumn{1}{c}{I/$\family$\,\%} & 
\multicolumn{1}{c@{\hskip 12pt}}{time} & 
1-by-1 & 
all-in-1\\\midrule


av-8-2      &  2e4     &  9e4     &  4e3         &  95.3    &  6.91    &  0.87      &  18.99   &  74&  2.7             &  \textbf{0.1}             \\
av-8-2-e    &  2e4     &  9e4     &  4e3         &  100   &  0.08    &  0.08      &  0.28    &  \textbf{3}&  51              &  1.2             \\
dodge-2     &  1e5     &  7e5     &  3e4         &  100   &  0.1     &  0.1       &  0.3     &  \textbf{124}&  5.3             &  1.4             \\
dodge-3     &  1e5     &  7e5     &  9e7         &  100   &  <0.01   &  <0.01     &  <0.01   &  \textbf{1453}&  $\dagger$1129   &  MO              \\
dpm-10      &  2e3     &  1e4     &  1e4         &  18.1    &  4.93    &  7.54      &  19.79   &  \textbf{20}&  1.8             &  32.4            \\
dpm-10-b    &  9e3     &  1e5     &  1e5         &  22.2    &  0.04    &  0.02      &  0.59    &  \textbf{78}&  18              &  MO              \\
obs-8-6     &  5e2     &  9e2     &  5e4         &  90.3    &  1.68    &  0.63      &  5.66    &  \textbf{7}&  3.4             &  1.1             \\
obs-10-6    &  8e2     &  3e3     &  2e6         &  98.4    &  <0.01   &  <0.01     &  0.04    &  \textbf{6}&  282             &  MO              \\
obs-10-9    &  1e3     &  2e3     &  4e8         &  100   &  <0.01   &  <0.01     &  <0.01   &  \textbf{275}&  $\dagger$1166   &  MO              \\
rov-100     &  2e3     &  5e4     &  2e7         &  46.9    &  0.24    &  <0.01     &  1.21    &  \textbf{728}&  $\dagger$52     &  MO              \\
rov-1000    &  2e4     &  5e5     &  4e6         &  99.8    &  0.34    &  0.03      &  1.2     &  \textbf{1441}&  $\dagger$51     &  MO              \\
uav-roz     &  2e4     &  2e5     &  4e3         &  98.9    &  1.62    &  0.02      &  10.09   &  \textbf{36}&  2.7             &  24.2            \\
uav-work    &  9e3     &  1e5     &  2e6         &  99.5    &  0.1     &  <0.01     &  0.39    &  \textbf{118}&  $\dagger$86     &  MO              \\
virus       &  2e3     &  1e5     &  7e4         &  83.3    &  14.31   &  0.85      &  49.92   &  413&  \textbf{0.7}             &  MO              \\
rocks-6-4   &  3e3     &  7e3     &  7e3         &  100   &  34.86   &  34.11     &  199.98  &  108&  0.2             &  \textbf{0.1}             \\
\bottomrule

\end{tabular}
}
\vspace{0.5em}
\caption{Information about the benchmarks and experimental synthesis results. The runtimes are in seconds. MO: $>$16GB. For the one-by-one evaluation, the runtimes exceeding the time limit of 2 hours (indicated by $\dagger$) are extrapolated. 
}
\vspace{-2.5em}
\label{tab:main}
\end{table}


\vspace{-.2em}
\subsection*{Q1 Efficacy and scalability}
Tab.~\ref{tab:main} reports the experimental results. We first report the model information discussed above. For our approach, including post-processing, we report four columns, where the first two concern the computed trees and the latter two the algorithm performance. In particular, they display:
(1)~The ratio L/$\family$ between the number of the tree leaves and the family size  $|\family|$. 
(2)~The ratio P/SAT between the number of leaves containing a policy and the number of satisfiable MDPs. 
(3) The ratio I/$\family$ between the total number of game \& MDP model checking calls required during the tree construction (from here on: the number of \emph{iterations})
and  $|\family|$. 
(4) The overall runtime. 
Finally, we provide the speedup obtained with respect to the baseline algorithms.


\paragraph{A small number of iterations.}
On the majority of benchmarks, our approach requires only a fraction (${<}$2\%) of iterations compared to the total number of MDPs, even for families containing millions of members.
The use of the game-based abstraction supplemented by the smart splitting heuristics allows our abstraction refinement approach to use (relatively) few splittings to obtain structurally similar sub-families.
First, this allows our approach to identify sub-families of unsatisfiable MDPs that are resolved by $\textsc{testUnsat}$.
Second, given a family of satisfiable MDPs, the splitting strategy also helps the game abstraction to find a few robust policies that together cover this family.

\paragraph{A challenging benchmark.} The \emph{rocks-6-4} model is the only benchmark where the ratio of iterations exceeds 100\%. In this benchmark,  the robot must deliver coffee to four desks in a room without obstacles. The location of the desks is parameterised. While there exists a single policy that is winning for all MDPs~--~a policy that visits each cell of the room~--~the proposed game abstraction cannot find it:
as long as the (sub-)family \family contains at least two MDPs, in the corresponding game abstraction, Player~2 may put the desk out of the way of the agent. This model clearly demonstrates the principal limitation of the game abstraction. A similar limitation can also be observed in the \emph{virus} model.


\vspace{-.1em}
\paragraph{Comparison with the baselines.}
Comparing the proposed method to the one-by-one enumeration, we observe speedups of orders of magnitude for many benchmarks. Typically, the acceleration increases when increasing the family size, see differences in the variants of ~\emph{dodge}, \emph{dpm} and \emph{obstacles} benchmarks. This is given by the fact that the ability of the game abstraction to reason about larger families is naturally more beneficial in the larger variants of these models. The largest speedup is observed on \emph{obs-10-6} model: the proposed method constructs a solution in under 5 minutes, whereas one-by-one enumeration would take at least 3 days. On the other hand, on the models where the abstraction is too coarse and splitting does not help (see above), the one-by-one enumeration outperforms our approach. In some models, the symbolic representation in the all-in-one baseline can capture similarities across the individual MDPs effectively; For such models, the speedup achieved by our approach is reduced compared to the one-by-one enumeration. However, for a large set of models, the symbolic method runs out of available memory when constructing the symbolic representation itself.



\begin{table}[t]
\renewcommand{\arraystretch}{0.95}
\setlength{\tabcolsep}{2pt}
\centering
\scalebox{0.95}{
\begin{tabular}{l@{\hskip 12pt}rrr@{\hskip 12pt}rrr@{\hskip 12pt}rr}
\toprule
\multicolumn{1}{c}{\multirow{2}{*}{model}} &
\multicolumn{3}{c@{\hskip 12pt}}{before} &
\multicolumn{3}{c@{\hskip 12pt}}{after} &
&
\multicolumn{1}{c}{time}
\\ \cmidrule(lr{1.5em}){2-4}\cmidrule(lr{1.5em}){5-7}

&
\multicolumn{1}{c}{nodes} &
\multicolumn{1}{c}{leaves} &
\multicolumn{1}{c@{\hskip 12pt}}{policies} &
\multicolumn{1}{c}{nodes} &
\multicolumn{1}{c}{leaves} &
\multicolumn{1}{c@{\hskip 12pt}}{policies} &
\multicolumn{1}{c}{time} &
\multicolumn{1}{c}{\%}
\\
\midrule

av-8-2      & 569 & 312 & 184 & 460 & 283 & 34 & $2$ & 2.94 \\ 
av-8-2-e    & 9 & 8 & 8 & 4 & 3 & 3 & ${<}1$ & 0 \\ 
dodge-2   & 76 & 61 & 61 & 44 & 29 & 29 & $11$ & 9.02 \\ 
dodge-3   & 796 & 637 & 637 & 375 & 246 & 246 & $184$ & 12.73 \\ 
dpm-10         & 1739 & 1159 & 314 & 1275 & 719 & 199 & ${<}1$ & 0 \\ 
dpm-10-b     & 546 & 363 & 300 & 89 & 54 & 5 & $1$ & 1.35 \\ 
obs-8-6  & 1848 & 1134 & 1055 & 1495 & 783 & 266 & ${<}1$ & 0 \\ 
obs-10-6 & 745 & 586 & 584 & 303 & 172 & 98 & ${<}1$ & 0 \\ 
obs-10-9 & 9721 & 6481 & 6481 & 7487 & 4247 & 2672 & $7$ & 2.7 \\ 
rov-100      & 2.0e5 & 1.0e5 & 3.8e4 & 9.4e4 & 5.8e4 & 69 & $51$ & 7.07 \\ 
rov-1000     & 3.4e4 & 1.7e4 & 8728 & 2.2e4 & 1.3e4 & 1274 & $40$ & 2.85 \\ 
uav-roz        & 381 & 259 & 234 & 124 & 73 & 1 & ${<}1$ & 0 \\ 
uav-work  & 5056 & 2562 & 1770 & 2731 & 1649 & 6 & $1$ & 0.88 \\ 
virus          & 2.9e4 & 1.5e4 & 1.5e4 & 1.7e4 & 1.0e4 & 515 & $34$ & 8.76 \\
rocks-6-4      & 9841 & 6561 & 6561 & 3483 & 2287 & 2238 & $13$ & 12.75 \\ 

\bottomrule
\end{tabular}
}
\vspace{0.5em}
\caption{Impact and cost of post-processing. Time is presented in seconds}
\label{tab:post-processing}
\vspace{-2em}
\end{table}

\subsection*{Q2: Size of the policy trees}
The size of the tree is directly related to the number of iterations the algorithm performs. Thus, we observe that 
for the majority of benchmarks, the constructed tree contains fewer leaves than 1\% of the total number of MDPs. This means that, on average, each leaf of the policy tree describes hundreds of MDPs for which either a robust policy was found, or no winning policy exists.
This ratio drops below 0.001\% for e.g.~\emph{dodge-3} benchmark, where we find 246 policies that satisfy 94 million MDPs.

\paragraph{Post-processing.} The proposed post-processing routine plays a prominent role in providing compact policy trees.
Tab.~\ref{tab:post-processing} presents an empirical evaluation of the proposed policy tree post-processing algorithm on all of the considered benchmarks. It reports the number of nodes, leaves and policies before and after the post-processing step as well as the total time post-processing takes in seconds and the portion of time relative to the total synthesis time. The post-processing indeed does not guarantee that the minimal number of policies is found (e.g. despite the threefold reduction for the \emph{rocks-6-4} model, we are still very far from the single policy that covers the entire family). On the other hand, for the majority of the models, it significantly reduces the size of the tree as well as the number of policies while introducing only a reasonable overhead -- typically only a few percent of the total synthesis time. In the \emph{rover-100} model, it achieves the largest reduction in the number of policies: from 38k to 69 with a 7\%~overhead.


\subsection*{Ablation study}
Finally, we consider variations of Alg.~\ref{alg:general} discussed in Sec.~\ref{sec:algorithm:randomization} and~\ref{sec:algorithm:splitting}, namely the use of randomised abstraction $\game^{\schedrandom}$ and alternative splitting strategies. Table~\ref{tab:variants} reports an evaluation of the default algorithm~--~using game abstraction~$\game$, splitting wrt.~game policy $\gsched$ and taking into account unreachable states~--~wrt. three alternatives: using randomised abstraction $\game^{\schedrandom}$, using pessimistic splitting $\qsched$ or optimistic splitting $\gsched$, both on a reachable fragment. We are interested in the number of nodes/policies in the tree, the number of iterations and the synthesis time. In each column, we report the ratio of the value achieved by the default method wrt.~the value achieved by the alternative method; values above~1 (typeset in bold) indicate that the corresponding alternative method is better than the default one.

\paragraph{Using randomised abstraction $\game^{\schedrandom}$ for finding robust policies.}
Our experiments confirm that $\game^{\schedrandom}$ is not competitive to the game abstraction: although solving $\game^{\schedrandom}$  is much faster than solving $\game$, the resulting policies are significantly less robust and thus the algorithm has to perform a larger number of iterations that leads to larger trees. Consequently, larger models are completely out of reach for the randomised abstraction. The small variant of the \emph{dpm-10} model is the only case where $\game^{\schedrandom}$ slightly outperforms $\game$ -- a detailed inspection shows that this is only due to a pessimistic splitting strategy that is used in $\game^{\schedrandom}$(see~below). 

\paragraph{Alternative splitting strategies.}
We investigate the impact of splitting using game policy $\gsched$ and taking into account unreachable states, further denoted as $\gsched^U$, compared to $\gsched$ or $\qsched$ (both on a reachable fragment). The pessimistic splitting wrt.~$\qsched$ is more effective on models with a low presence of satisfiable MDPs (e.g., the \emph{dpm-10} models with around 20\% SAT) and typically achieves a mild improvement (below a factor 2) compared to $\gsched^U$.  
Remarkably, on the \emph{rover} models, the improvement is significant: e.g. a 15-fold time improvement on the \emph{rover-100} model (47\% SAT). On the other hand, the pessimistic splitting can be very inefficient on large families where all MDPs are satisfiable: it reached three timeouts in our benchmark suite. The reachable restriction $\gsched$ mildly lags behind $\gsched^U$ on a majority of the models with two exceptions: on the \emph{rover-100} model, it improves the runtime by a factor of 2.6, but on the \emph{dpm-10-big} model, it timeouts due to a large number of unsatisfiable MDPs. Overall, the default method that uses $\gsched^U$ offers the most stable and scalable performance.

\begin{table}[h]
\renewcommand{\arraystretch}{0.95}
\setlength{\tabcolsep}{1pt}

\scalebox{0.95}{
\begin{tabular}{l@{\hskip 12pt}cccc@{\hskip 12pt}cccc@{\hskip 12pt}cccc}

\toprule
\multicolumn{1}{c}{\multirow{2}{*}{model}} &
\multicolumn{4}{c@{\hskip 12pt}}{using $\game^{\schedrandom}$} &
\multicolumn{4}{c@{\hskip 12pt}}{using $\gsched$} &
\multicolumn{4}{c}{using $\qsched$}
\\
\cmidrule(lr{1.7em}){2-5}\cmidrule(lr{1.7em}){6-9}\cmidrule(lr{0.75em}){10-13}

& 
\multicolumn{1}{c}{nodes} &
\multicolumn{1}{c}{polic.} &
\multicolumn{1}{c}{iters} &
\multicolumn{1}{c@{\hskip 12pt}}{time} &
\multicolumn{1}{c}{nodes} &
\multicolumn{1}{c}{polic.} &
\multicolumn{1}{c}{iters} &
\multicolumn{1}{c@{\hskip 12pt}}{time} &
\multicolumn{1}{c}{nodes} &
\multicolumn{1}{c}{polic.} &
\multicolumn{1}{c}{iters} &
\multicolumn{1}{c}{time}
\\ \midrule

av-8-2      & 0.97 &     0.81 &     0.23 &     0.18 & \textbf{1.01} &     1 &    0.88 &     0.78 & \textbf{1.06} &     \textbf{1.7} &  0.93 &     0.8 \\
av-8-2-e    & 0.1 &  0.1 &  0.01 &     0.02 & 1 &    1 &    1 &    1 & 1 &    1 &    2 &    0.75 \\ 
dodge-2   & 0.46 &     0.43 &     0.22 &     0.41 & 1 &    1 &    1 &    0.92 & 0.79 &     0.81 &     0.42 &     0.52 \\ 
dodge-3   & \multicolumn{4}{c}{TO} & \textbf{1.01} &     \textbf{1.01} &     0.99 &     0.97 &\multicolumn{4}{c}{TO} \\
dpm-10         & \textbf{1.89} &     \textbf{1.12} &     \textbf{1.5} &  \textbf{1.8} & 0.3 &  0.42 &     0.37 &     0.36 & \textbf{1.75} &     \textbf{1.57} &     \textbf{2.11} &     \textbf{1.64} \\ 
dpm-10-b     & 0.11 &     0.02 &     0.29 &     0.73 & \multicolumn{4}{c}{TO} & 0.67 &     0.83 &     \textbf{2.85} &     \textbf{1.21} \\ 
obs-8-6  & \textbf{1.21} &     0.6 &  0.81 &     \textbf{1.5} & 0.77 &     0.42 &     0.72 &     0.75 & 0.16 &     0.08 &     0.19 &     0.24 \\ 
obs-10-6 & 0.01 &     0.01 &     0.01 &     0.01 & 0.77 &     0.49 &     \textbf{1.19} &     \textbf{1.25} &\multicolumn{4}{c}{TO} \\
obs-10-9 & \multicolumn{4}{c}{TO} & \textbf{1.18} &     0.84 &     \textbf{1.22} &     \textbf{1.27} &  \multicolumn{4}{c}{TO} \\ 
rov-100      & \textbf{1.39} &     0.12 &     0.85 &     0.49 & \textbf{2.59} &     \textbf{1.64} &     \textbf{2.01} &     \textbf{2.59} & \textbf{20.8} &    0.85 &     \textbf{18.95} &    \textbf{14.71} \\ 
rov-1000     &  \multicolumn{4}{c}{TO} & \textbf{1.08} &     0.76 &     0.98 &     0.97 & \textbf{9.81} &  \textbf{3.6} &  \textbf{8.27} &     \textbf{5.63} \\ 
uav-roz        & 0.9 &  0.25 &     0.49 &     \textbf{1.4} & 1 &    1 &    0.92 &     \textbf{1.03} & 0.93 &     1 &    0.81 &     0.73 \\ 
uav-work   & 0.99 &     0.86 &     0.46 &     0.81 & 1 &    1 &    0.77 &     0.61 & 1 &    1 &    0.77 &     0.61 \\ 
virus          & 0.91 &     0.86 &     0.53 &     0.46 & 0.77 &     0.94 &     0.75 &     0.78 & 0.72 &     0.69 &     0.65 &     0.8 \\ 
rocks-6-4    & 0.87 &     0.9 &  0.8 &  0.77 & 0.69 &     0.74 &     0.68 &     0.55 & 0.84 &     0.87 &     \textbf{1.01} &     0.86 \\ \bottomrule

\end{tabular}
}
\vspace{.5em}
\caption{Evaluation of alternative variants. TO indicates reaching the time limit of 1 hour. The columns represent a relative comparison with the default algorithm.}
\label{tab:variants}
\vspace{-2em}
\end{table}

The ablation study shows that solving $\game^{\schedrandom}$ results in significantly less robust policies leading to larger number of iterations. We also show that pessimistic splitting can be very inefficient on families where all MDPs are satisfiable.


\section{Related Work}
\label{sec:related}
Synthesis of robust policies as well as the verification of sets of Markov models are broad topics. This paper brings them together using a game-based approach. Below, we discuss these (overlapping) areas as well as closely related aspects such as permissive policies and compact policy representation.

\paragraph{Game-based abstraction.}
Game-based abstraction was primarily explored towards the verification of very large or infinite MDPs~\cite{DBLP:conf/tacas/HahnHWZ10,DBLP:journals/fmsd/KattenbeltKNP10}. In that setting, Player 1 vertices correspond to abstract states that partition the state space of the original MDP: Player 1 thus chooses a concrete state and Player 2 chooses an available action from this state. The literature focuses on extracting predicates 
allowing for a good partitioning. Recall that we use a different game construction, allowing us to reason about robust policies for a set of MDPs.
Game-based abstraction on POMDPs~\cite{DBLP:journals/tac/WintererJWJTKB21} is conceptually close to our game. However, for POMDPs, the abstraction cannot use the specific structure of our problem, and for refinement towards a robust policy, it can (and must) add memory to the policies, whereas we aim to find a concise policy tree via divide-and-conquer.


\paragraph{Families of MCs.}
Feature-based models are popular to describe software product lines~\cite{onebyone,lanna2018feature} and are conceptually similar to our quotient MDPs. 
Closest is work on abstraction-refinement on the quotient MDP to partition a family of MCs based on a specification~\cite{cegar,andriushchenko2021inductive} and alternative approaches that prune these families based on counterexamples obtained on a single model~\cite{cegis-journal}. Neither supports nondeterminism, thus no policies can be obtained.
The closest all-in-one approach is presented in~\cite{DBLP:journals/fac/ChrszonDKB18} and also discusses how to represent the optimal value for every MDP. Symbolic all-in-one approaches profit from variable reordering~\cite{DBLP:journals/corr/abs-2004-13287}.

\paragraph{Parametric MDPs.}
Parametric Markov models use constant but unknown transition probabilities. Families can be seen as a special case, a precise relation is discussed in \cite[Ch.\ 6]{DBLP:phd/dnb/Junges20}. Algorithms for parametric MDPs, however, require that the parameter choices do not change the underlying graph. Under that assumption, both formalisms are orthogonal. 
For parametric MDPs, a game-based approach is known as \emph{parameter lifting} technique~\cite{DBLP:journals/corr/abs-1903-07993}. It creates an SG abstraction, but the policies themselves are not extracted there, and no refinement techniques are studied. There is limited work on robust memoryless policies in pMDPs~\cite{DBLP:conf/concur/WinklerJPK19}: They are complexity-wise harder to obtain than in our setting, unless so-called rectangularity is assumed, which leads to variations of \emph{interval MDPs}~\cite{DBLP:journals/ai/GivanLD00,DBLP:journals/ior/NilimG05,DBLP:journals/mor/Iyengar05,DBLP:conf/cav/PuggelliLSS13}. In that context, a game-based view on robustness is popular~\cite{DBLP:journals/mor/Iyengar05,DBLP:conf/cdc/WeiningerMK19}.
The existence of small policy sets over pMDPs has been studied (theoretically)   in~\cite{DBLP:conf/concur/WinklerJPK19} and empirically as a by-product of a verification routine~\cite{DBLP:journals/sttt/HahnHZ11}. 

\paragraph{Robust policies in hidden-model MDPs.}
Finding one policy for a set of MDPs has been studied in the context of sets of MDPs (under different names and with different assumptions).
 Algorithmically, a robust policy can then be found by mapping this to POMDP synthesis. For MDPs with parametric transition probabilities, this has been studied in~\cite{arming2018parameter} (supporting up to 10 MDPs or 300 states).
The work in  \cite{DBLP:conf/aips/ChatterjeeCK0R20} considers specializations of POMDP planners (both offline and online) suitable for discounted expected rewards. The experiments cover benchmarks with tens of MDPs. The work in \cite{DBLP:conf/aaai/ChadesCMNSB12} considers a specialization of point-based solvers with a particular interest in obtaining compactly represented robust policies.
Qualitative properties are studied in \cite{DBLP:conf/fsttcs/RaskinS14,DBLP:conf/tacas/VegtJJ23} (by considering policies with memory). Alternatively, assuming a prior on the set of MDPs, one can optimize the expected performance~\cite{DBLP:journals/mmor/BuchholzS19,DBLP:journals/iiset/SteimleKD21}. Finally, as an alternative, PAC-robustness over a distribution over sets of MDPs has been studied~\cite{DBLP:journals/sttt/BadingsCJJKT22}.

\paragraph{Data-driven robust policies.}
In the setting where the transition and reward functions are unknown, reinforcement learning (RL) is applicable. Robustness in RL is studied in its relation to \emph{generalisation}, see~\cite{DBLP:conf/nips/AzarLB13,DBLP:journals/corr/abs-2111-09794}. The field of curriculum learning~\cite{DBLP:journals/jmlr/NarvekarPLSTS20} considers learning policies for different but related tasks, motivated to ease the learning of complex tasks.

\paragraph{Permissive policies and shielding.}
An alternative approach towards finding robust memoryless policies is to compute all (memoryless) winning policies for every MDP and to construct their intersection. The sets of policies can be captured via permissive policies that allow for a nondeterministic action choice~\cite{DBLP:journals/corr/DragerFK0U15} and are relevant in the context of shielding for MDPs~\cite{DBLP:journals/fmsd/KonighoferABHKT17,DBLP:conf/tacas/Junges0DTK16}. However, preliminary results in this direction confirmed that the computational overhead of computing permissive policies on individual MDPs is prohibitive.

\vspace{-.1em}
\paragraph{Compact policies.}
To compactly represent individual policies, decision trees are a popular target, either via a direct search~\cite{DBLP:conf/ijcai/VosV23} or as post-processing~\cite{DBLP:conf/tacas/AshokJKWWY21}. Concise representations for policies with memory can be synthesised with techniques akin to this paper~\cite{andriushchenko2022inductive} or as a post-processing step~\cite{borktacas24}.
Finally, such procedures can also be realised by first learning a (recurrent) neural network and then extracting an automaton from this net~\cite{DBLP:journals/jair/Carr0T21}.

\vspace{-1em}
\section{Conclusion and Future Work}
\vspace{-0.5em}

We have presented a divide-and-conquer algorithm enabling the construction of policy trees that compactly represent winning policies in large families of MDPs. The algorithm leverages a novel form of a game abstraction supplemented by smart abstraction-refinement strategies. Our extensive experimental evaluation demonstrates superior scalability compared to existing baselines as well
as novel alternatives. For future work, we are interested in i) further investigating strategies to search for robust policies, potentially incorporating richer feedback from model checking subroutines similar to inductive synthesis approaches~\cite{andriushchenko2021inductive} and ii)~combining a compact representation of policies (e.g. using decision diagrams) with the compact representation of the policy trees. 

\bibliographystyle{splncs04}
\bibliography{bibliography}

\clearpage

\appendix

\section{MDP Program Sketching}
\label{app:sketching}
Our approach assumes that there exists a compact representation of the family~\family. 
In our implementation and the experimental evaluation, we consider families represented by MDP \emph{sketches}. A sketch~\cite{DBLP:journals/sttt/Solar-Lezama13} is a syntactic template defining a high-level structure of an underlying operational model. It allows for a concise representation of a priori knowledge about the system under development. Sketches for finite-state probabilistic programs (i.e. finite-state MCs) written in the Prism guarded-command language~\cite{KNP11} have been proposed in~\cite{cegis-journal}. The sketch contains a finite number of \emph{holes} representing the program’s open, i.e., undefined parts. A hole may be used in the guard or updates of a command similar to a constant and
may occur multiple times within one or multiple commands. Each hole is associated with a finite domain representing the possible options in which the hole can be filled. In this work, we extend the MC sketches to MDP sketches. Since the extension is straightforward, we omit the formal definition and rather illustrate MDP sketching by presenting the sketch for our running example in Fig.~\ref{fig:running-example-sketch}.

\lstset{language=Prism}   



\newsavebox{\sketchbox}
\begin{lrbox}{\sketchbox}
\begin{lstlisting}[numbers=none]
mdp

hole int OX in {2..5};                 module clock     
hole int OY in {2..4};                   clk : [0..1] init 0;
formula goal = (x=6 & y=6);              [l] !done & clk=0 -> (clk'=1);
formula done = goal | crash;             [r] !done & clk=0 -> (clk'=1);
formula xr = min(x+1,6);                 [d] !done & clk=0 -> (clk'=1);
formula xl = max(x-1,1);                 [u] !done & clk=0 -> (clk'=1);
formula yu = min(y+1,6);                 [crash] !done & clk=1 -> (clk'=0);
formula yd = max(y-1,1);               endmodule

module agent
  x : [1..6] init 1;
  y : [1..6] init 1;
  crash : bool init false;
  [l] true -> 0.91: (x'=xl) + 0.03: (y'=yd) + 0.03: (y'=yu) + 0.03: (x'=xr);
  [r] true -> 0.91: (x'=xr) + 0.03: (y'=yu) + 0.03: (y'=yd) + 0.03: (x'=xl);
  [d] true -> 0.91: (y'=yd) + 0.03: (x'=xr) + 0.03: (x'=xl) + 0.03: (y'=yu);
  [u] true -> 0.91: (y'=yu) + 0.03: (x'=xl) + 0.03: (x'=xr) + 0.03: (y'=yd);
  [crash] true -> (crash'=(x=OX & y=OY)); //instantiation affects this line
endmodule
\end{lstlisting}
\end{lrbox}

\newsavebox{\instancebox}
\begin{lrbox}{\instancebox}
\begin{lstlisting}[numbers=none]
module rex
s : [0..3] init 0;
s = 0 -> 0.5: s'=1 + 0.5: s'=3;     
s = 1 -> 1: s'=s+2; 
s >= 2 -> 1: s'=s;
endmodule
\end{lstlisting}
\end{lrbox}

\begin{figure}[t]
    \centering
    \usebox{\sketchbox}
    \vspace{-0.5em}
    \caption{Running example from Fig.~\ref{fig:running-example:grid} encoded as a Prism program sketch. ``Hole" variables OX and OY can be instantiated to one of the values in their specified domain, and when all holes are instantiated, we obtain one concrete MDP.\vspace{-1.5em}}
    \label{fig:running-example-sketch}
\end{figure}

\vspace{-1em}
\section{Proofs}
\vspace{-.5em}
\label{app:proofs}

This section provides proof of the Theorems and Lemmas presented in our paper.

\subsubsection{Proof of Lemma~\ref{lemma:quotient-maxmax}}
Assume $\probmax[\qmdp_\family]{\F T} < \lambda$ and that there exists MDP $M_i$ s.t.~$\exists \sigma \in \Sigma^M \ \colon \reach{M_i^\sigma} \geq \lambda$.
Construct policy $\overline{\sched}$ as follows: $\overline{\sched}(s) = (\sched(s),I)$ where $I \in \indicespartition$ s.t. $i \in I$ and $\act = \sched(s)$.
Note that $\qmdp_\family^{\overline{\sched}}$ is the same MC as $M_i^{\sched}$: in each state $s \in \states$ of the quotient, policy $\overline{\sched}$ executes action $\sched(s)$ from MDP $M_i$.
Thus, $\reach{\qmdp_\family^{\overline{\sched}}} = \reach{M_i^\sigma} \geq \lambda$, i.e.~policy $\overline{\sched}$ achieves in MDP $\qmdp$ value larger than $\probmax[\qmdp_\family]{\F T}$, which is a contradiction.

\subsubsection{Proof of Theorem~\ref{theorem:policy}} If $\playeronemax[\game_{\family}]{\F{T}} \geq \lambda$ under the optimal policy $\sigma = (\sched[1], \sched[2])$, then from the definition of $\playeronemax[\game_\family]{\F{T}}$ we know that: 
$$\forall \sched[2]' \in \schedulers_2: \reach{\game_{\family}^{\sched[1],\sched[2]'}} \geq \reach{\game_{\family}^{\sched[1],\sched[2]}} \geq \lambda.$$
This means that no matter what strategy Player 2 chooses, the value of the game abstraction will stay $\geq \lambda$. Assume arbitrary $i \in \indices$ and let $\sched[2]' \in \schedulers_2$ be a policy that is consistent in identifier $i$. From Lemma~\ref{lemma:player2-consistency} it follows that:
$$\reach{M_i^{\sched[1]}} = \reach{\game_{\family}^{\sched[1],\sched[2]'}} \geq \lambda$$
and therefore $\sched[1]$ is a robust policy for family $\family$.

\subsubsection{Proof of Theorem~\ref{theorem:correctness}}
First, we remark that if the subfamily is a singleton, there is either a winning policy for this MDP (which is then a robust policy for that subfamily), or the MDP is not winning, and thus all MDPs in the subfamily are unsatisfiable. The recursion must be finite, and the algorithm must terminate since we consider finite families only and \textsc{createSplit} returns non-trivial subsets. In fact, because of this, the recursion is linearly bounded by the number of MDPs in the family. The output is a valid policy tree following Theorem~\ref{theorem:policy} (covering correctness of \textsc{findRobust}) and Lemma~\ref{lemma:quotient-maxmax} (covering correctness of \textsc{testUnsat}).

\newpage

\section{Policy Tree Example}
\vspace{-1em}
\label{app:tree}

\begin{figure}[h]
\centering
    \includegraphics[width=0.9\textwidth]{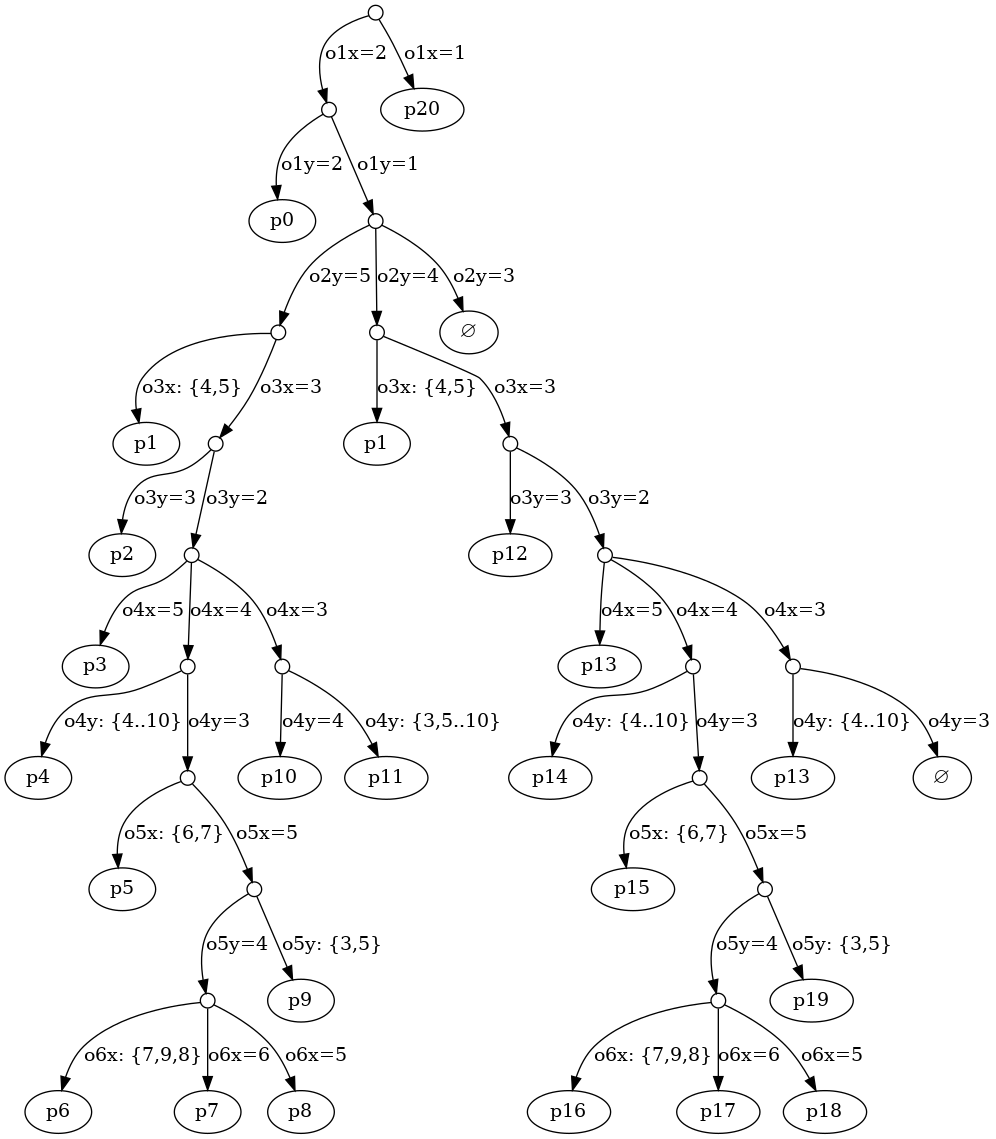}
    \label{fig:policy-tree-example}
    \caption{Policy tree generated for a sub-family of \emph{obstacles-10-6} wher $o2x{=}1$ and $o2y{\in}\{3..5\}$.}
\end{figure}

\end{document}